\documentclass[epj]{svjour}
\usepackage{amsmath,amssymb,psfig,epsfig,wrapfig}
\def\beq{\begin{equation}}
\def\eeq{\end{equation}}
\def\beqa{\begin{eqnarray}}
\def\eeqa{\end{eqnarray}}
\begin{document}
\title{Rescattering effects in hadron-nucleus and heavy-ion collisions}
\author{Jianwei Qiu\inst{1}
}                     
\institute{Department of Physics and Astronomy, 
           Iowa State University,
           Ames, Iowa 50011, U.S.A.}
\date{Received: \today}
\abstract{
We review the extension of the factorization formalism
of perturbative QCD to {\it coherent} soft rescattering associated
with hard scattering in high energy nuclear collisions.
We emphasize the ability to quantify high order corrections 
and the predictive power of factorization approach 
in terms of universal nonperturbative matrix elements.  
Although coherent rescattering effects are power suppressed 
by hard scales of the scattering, 
they are enhanced by the nuclear size and 
could play an important role in understanding 
the novel nuclear dependence observed
in high energy nuclear collisions.
} 
\maketitle
%
\section{Introduction}
\label{intro}

Rescattering in hadron-nucleus and heavy-ion collisions provides an
excellent tool to diagnose properties of nuclear medium and could
play an important role in understanding novel nuclear dependence
recently observed at relativistic heavy ion collider (RHIC) and in
planning future experiments at the Large Hadron Collider
(LHC). Many approaches in studying the rescattering effects have been
proposed and used for calculating nuclear dependence in high energy
nuclear collisions 
\cite{YellowReport,Qiu:2001hj,Qiu:2004id,BDMPS,Zakharov,Gyulassy,GW-loss}.

In this talk, we focus on a treatment of {\it coherent} soft
rescattering associated with hard probes \cite{Qiu:2001hj,Qiu:2004id}.
Our work is based on perturbative QCD (pQCD) factorization approach,
which is different from the works of Baier {\it et al.} (BDMPS)
\cite{BDMPS} and Zakharov \cite{Zakharov}, and the reaction operator
approach of Gyulassy {\it et al.} \cite{Gyulassy}.  The BDMPS analysis
does not require the presence of a hard scattering, but describes the
coherent results of many soft scatterings.  Its primary subject is
induced energy loss.  Our analysis requires a hard scale, and begins
with the pQCD treatment of hard-scattering with emphases on
momentum transfer, caused by coherent initial- and final-state soft
scatterings \cite{Qiu:2001hj,Qiu:2004id}.  
Our work attempts to stay as close as
possible to the pQCD factorization formalism, in which we may readily
quantify high order corrections in powers of strong coupling constant
$\alpha_s$, as well as corrections that decrease with extra powers of
momentum transfer 
\cite{Qiu:1990xy,Qiu:2002mh,Qiu:2003cg,Qiu:2003vd}.  

In the following we consider only initial- and final-state interaction
that gives leading power in medium length $(A^{1/3})$ and in
$\alpha_s$ at each scattering.  We first identify the coherence length
in nuclear collisions and the source of the leading medium size
enhancement. We then apply pQCD factorization approach to calculate
the leading nuclear dependence in several physical observables.  We
show that if we neglect soft rescattering off quark fields 
\cite{Jaffe:1981td,Qiu:1990xx}, the
leading medium effects induced by multiple soft rescattering depend on
only one well-defined nonperturbative matrix element,  
$\langle F^{+\alpha}F^{\ +}_{\alpha}\rangle$, defined below.   
We extract its value from different physical measurements and discuss
its universality.  A brief summary is given at the end. 

\begin{figure}
\begin{center}
  \epsfig{figure=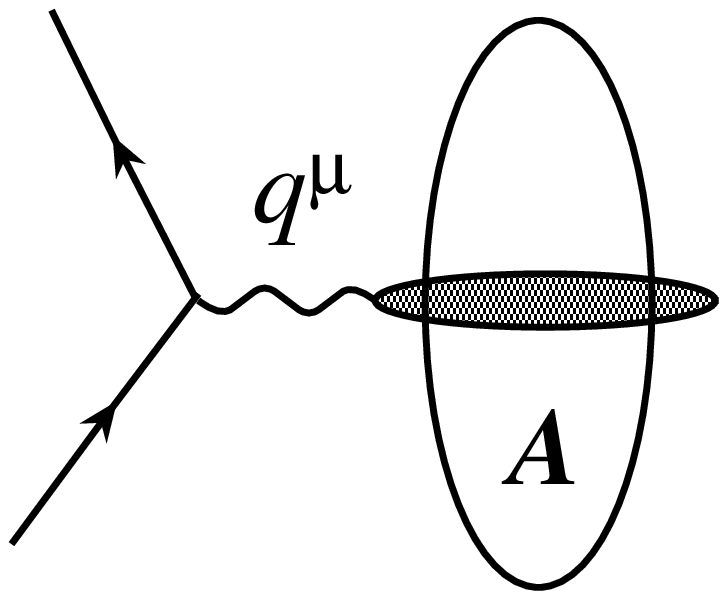,width=0.11\textwidth}
  \hskip 0.1in
  \epsfig{figure=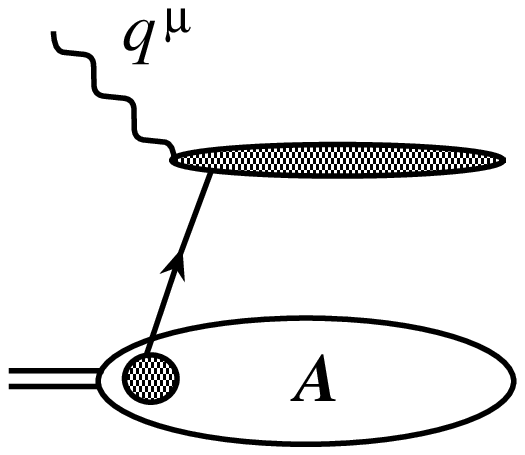,width=0.105\textwidth}
  \hskip 0.05in
  \epsfig{figure=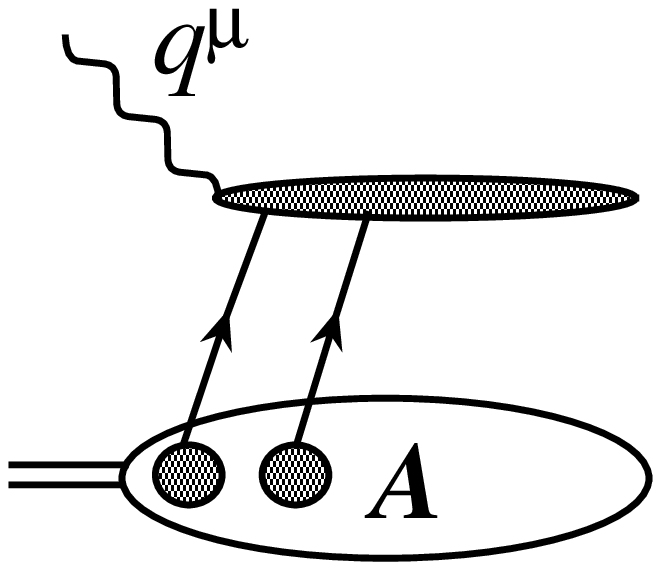,width=0.105\textwidth}
  \hskip 0.05in
  \epsfig{figure=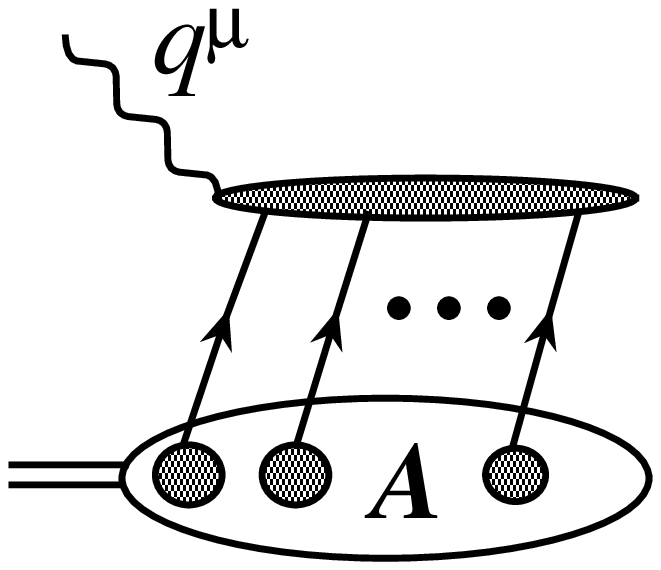,width=0.105\textwidth}

  (a)
  \hskip 0.7in (b)
  \hskip 0.6in (c)
  \hskip 0.6in (d)
\caption{Sketch for the scattering of a parton (or a lepton) in a
  large nucleus with a momentum transfer $q^\mu$.}
\label{fig1}
\end{center}
\end{figure}

\section{Coherent multiple scattering and leading nuclear 
         $A$-dependence}
\label{sec:2}

A hard probe corresponds to a scattering process with a large momentum
exchange $q^\mu$ whose invariant mass 
$Q\equiv \sqrt{|q^2|} \gg \Lambda_{\rm QCD}$, as sketched in
Fig.~\ref{fig1}.  It can probe a 
space-time dimension much smaller than a nucleon {\it at rest}, 
$1/Q\ll 2R \sim $~fm, with nucleon radius $R$.  But, the same probe
might cover a whole Lorentz contracted large nucleus, if $1/Q>2R(m/p)$
with averaged nucleon momentum $p$ and mass $m$, or equivalently,
$x\ll x_c \equiv 1/2mR\sim 0.1$ with $x$ being an active parton
momentum fraction in the scattering, $xp\sim Q$.  
The critical value $x_c$ corresponds to the nucleon size.  
If the active $x$ is much smaller than $x_c$, a hard
probe could cover several nucleons in a Lorentz contracted large
nucleus and interact with partons from different nucleons {\it
coherently}.  

\begin{figure}
\begin{center}
  \epsfig{figure=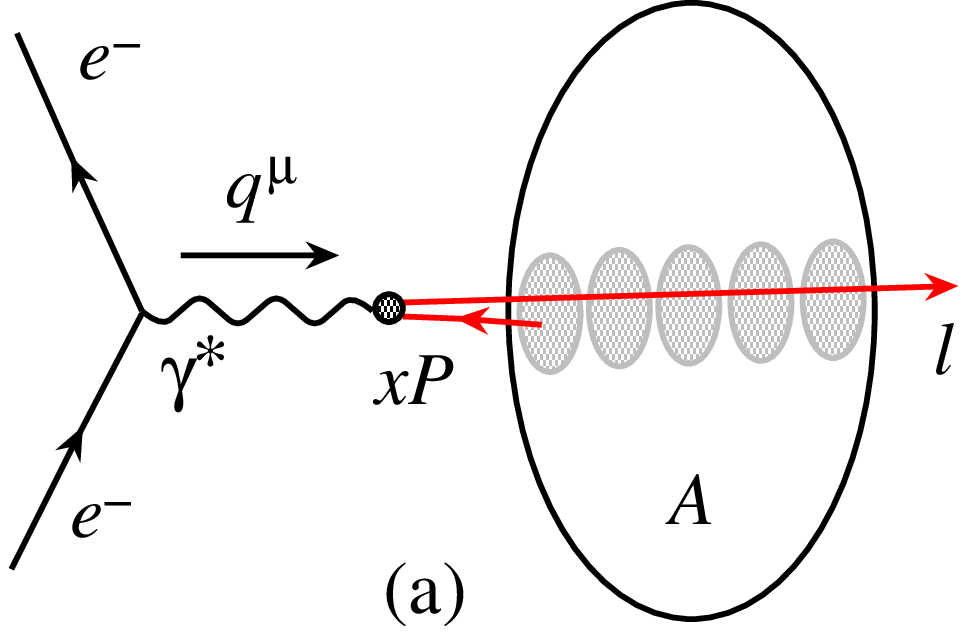,width=0.20\textwidth}
  \epsfig{figure=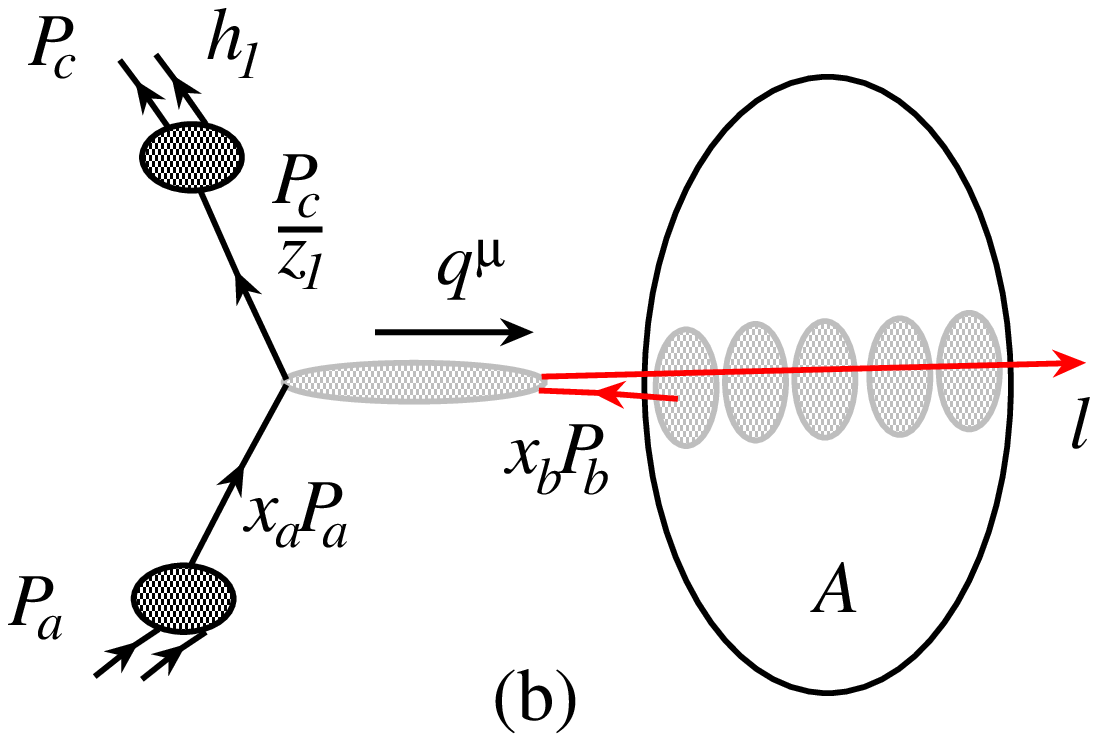,width=0.20\textwidth}
\caption{Coherent multiple scattering of the struck parton in deeply
  inelastic scattering (a) and in $t$-channel in hadron-nucleus
  collisions (b).}
\label{fig2}
\end{center}
\end{figure}

Inclusive deeply inelastic scattering (DIS) on a nucleus offers an
ideal example of coherent multiple scattering and power corrections
\cite{Qiu:2003vd}.  The strength of the scattering is defined by the
virtual photon momentum, $q^\mu$.
Let $q^\mu=-x_B p^\mu + Q^2/(2x_B p\cdot n)n^\mu$ with $n^\mu$ defined
along a lightcone direction opposite to that of $p^\mu$.   
While the $Q^2=-q^2$ sets up the hard collision scale, the scattered
quark probes the nuclear matter via multiple soft final-state
interactions.  When Bjorken $x_B\equiv Q^2/2p\cdot q \ll x_c$, 
the multiple scatterings at a given impact parameter are coherent over
entire size of the Lorentz contracted nucleus.  
This is best seen in the Breit frame,
as shown in Fig.~\ref{fig2}(a), where the incoming quark reverses its
direction after interacting with the virtual photon and collides with
the ``remnants'' of the nucleus at the same impact parameter.  
The same coherent multiple scatterings can take place in
hadron-nucleus collisions along the direction of momentum exchange of
the scattering \cite{Qiu:2004da}, as shown in Fig.~\ref{fig2}(b).

To identify the leading medium length enhanced nuclear effects, let's
consider multiple scattering contribution to inclusive DIS off a large  
nucleus, as sketched in Fig.~\ref{fig2}(a).  
For a spin-averaged inclusive DIS cross section, 
there is only one large momentum scale, $Q$,
``invariant mass'' of the exchanged virtual photon, and the
factorization is expected to be valid for all power corrections as a
consequence of operator product expansion (OPE) \cite{OPE}:
\beqa
d\sigma^{\gamma^* A}_{\rm DIS}
&=&
d\hat{\sigma}^{i}_{2} \otimes
\left[ 1 + c^{(1,2)}\alpha_s + c^{(2,2)}\alpha_s^2 + \dots \right]
\otimes T_{2}^{i/A}
\nonumber \\
&+&
\frac{d\hat{\sigma}^{i}_{4}}{Q^2} \otimes
\left[ 1 + c^{(1,4)}\alpha_s + c^{(2,4)}\alpha_s^2 + \dots \right]
\otimes T_{4}^{i/A}
\nonumber \\
&+&
\frac{d\hat{\sigma}^{i}_{6}}{Q^4} \otimes
\left[ 1 + c^{(1,6)}\alpha_s + c^{(2,6)}\alpha_s^2 + \dots \right]
\otimes T_{6}^{i/A}
\nonumber \\
&+&
\dots
\label{facdis}
\eeqa
where $\otimes$ represents convolutions in fractional momenta
carried by partons and $T_n$ represents a parton correlation function
or a matrix element of a twist $n$ operator.  In Eq.~(\ref{facdis}),
the $\hat{\sigma}^{i}_{n}$ and $c^{(j,n)}$ are perturbatively
calculable short-distance partonic cross sections or coefficient
functions, which are independent of the target size.  Therefore, we
need to find the nuclear size ($A^{1/3}$-type) enhancement 
induced by multiple rescattering from the matrix elements, if there is
any. 

\begin{figure}
\begin{center}
\epsfig{figure=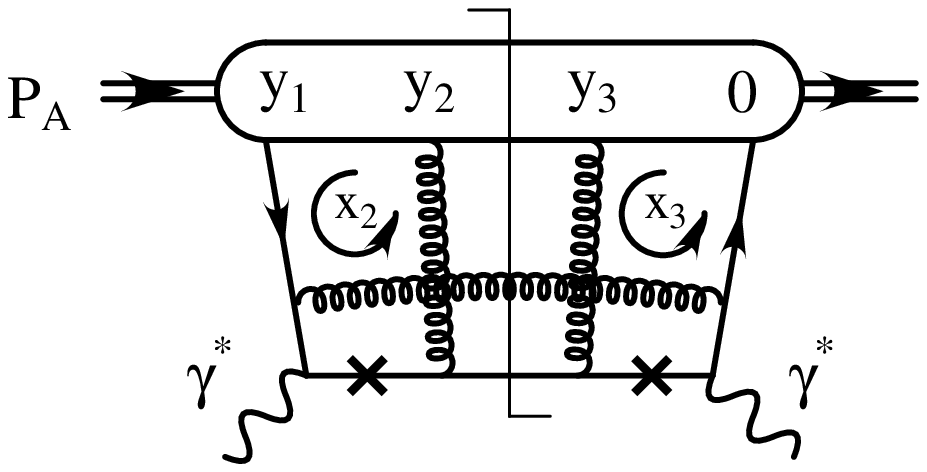,width=1.7in}
\hskip 0.1in
\epsfig{figure=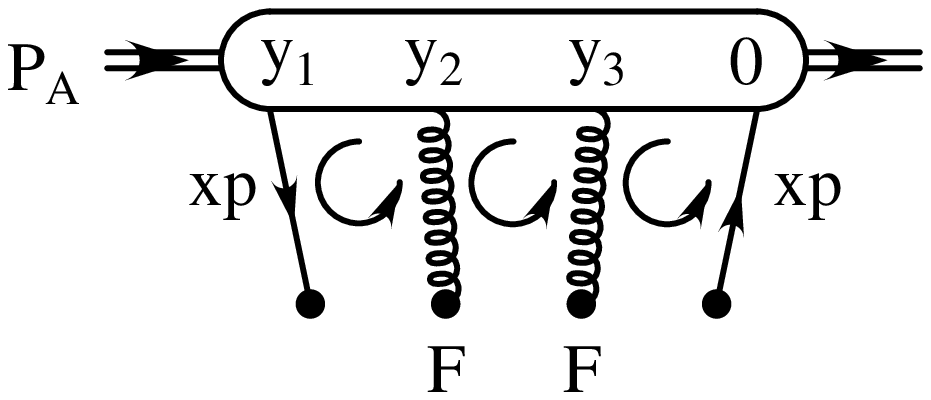,width=1.6in}

  (a) \hskip 1.4in (b)
\end{center}
\caption{(a) Diagram with poles that give rise to an $A^{1/3}$-type
  enhancement to DIS cross section off a large nucleus; and (b)
  corresponding twist-4 quark-gluon correlation function. } 
\label{fig3}
\end{figure}

The leading twist parton distributions, $T_{2}$'s, represent
probability densities to find a single parton in a target and can have
some nuclear dependence via the input distributions to
their DGLAP evolution equations.  The $A^{1/3}$-type target-size
enhancement to DIS cross section can only appear in the terms beyond
the first row in Eq.~(\ref{facdis}).  For definiteness, 
let's consider a leading power suppressed contribution to DIS cross
section, as shown in Fig.~\ref{fig3}(a),
which can be factorized into the form according to Eq.~(\ref{facdis})
\cite{Qiu:2001hj}, 
\beqa
d\sigma^{(4)}_A
&=&
\sum_{(ii')}
\int dx_1\, dx_2\, dx_3\; T^{(4)}_{(ii')/p_A}(x_1,x_2,x_3)\,
\nonumber \\
&\ & \times \ d\hat{\sigma}^{(4)}_{(ii')}(x_1,x_2,x_3)\, .
\label{sig4}
\eeqa
The matrix element $T^{(4)}$, as sketched in Fig.~\ref{fig3}(b), 
is typically of the form \cite{Qiu:1990xy},
\beqa
&&
T_{(ii')/p_A}(x_1,x_2,x_3)
\nonumber \\
&&\hskip 0.2in  \propto \int{dy^-_1dy^-_2dy^-_3\over(2\pi)^3}
{\rm e}^{ip^+(x_1y^-_1+x_2y^-_2+x_3y^-_3)}
\nonumber \\
&& \hskip 0.35in \times \, 
\langle p_A|
B^\dagger_{i}(0)B^\dagger_{i'}(y^-_3)B_{i'}(y^-_2)B_{i}(y^-_1)
|p_A\rangle\, ,
\label{Tdef}
\eeqa
where $p=p_A/A$ and $B_i$ is the field corresponding to a parton of
type $i=q,{\bar q},G$.  The structure of the target is manifest only
in the matrix element $T$ in Eq.~(\ref{sig4}). 
Each pair of fields in the matrix element Eq.~(\ref{Tdef}) represents
a parton that participates in the hard scattering.
The $y^-_i$ integrals cover the distance between
the positions of these particles along the path of the
outgoing scattered quark.
In Eq.~(\ref{Tdef}), integrals over the
distances $y^-_i$ generally cannot grow with the size
of the target because of oscillations from the exponential factors
${\rm e}^{ip^+x_iy^-_i}$.

Since the kinematics of a single-scale hard collision is only
sensitive to the total momentum from the target, two of the three
momentum fractions: $x_1,x_2$, and $x_3$ cannot be fixed by the
hard collisions.  Therefore,
there is always a subset of Feynman diagrams, 
like one in Fig.~\ref{fig3}(a) with poles labeled by the crosses 
``$\times$'', whose contribution to the partonic parts, 
$\hat{\sigma}_{(ii')}^{(4)}$ in Eq.~(\ref{sig4}) is dominated by
regions where two of the three momentum fractions vanish.
The convolution over $dx_1\,dx_2\,dx_3$ in Eq.~(\ref{sig4}) is
simplified to an integration over only one momentum fraction,
\beqa
& &
\int \prod_{i=1}^{3}\, dx_i\ T_{(jj')/p_A}(\{x_i\})\,
        \hat{\sigma}^{(4)}_{(jj')}(\{x_i\} p)
\nonumber \\
& & \Longrightarrow
\int dx\, T_{q}(x,A)\, \hat{\sigma}^{(D)}_{(q)}(x p),
\label{poles2x}
\eeqa
where the partonic part $\hat{\sigma}^{(D)}$ is finite and
perturbative, with the superscript $(D)$ indicating the contribution
from double scattering.  The above matrix element, $T_{q}(x,A)$, as
illustrated in Fig.~\ref{fig3}(b), has the form
\begin{eqnarray}
T_{q}(x,A) 
&=& 
\int {d y^-_1 \over 2\pi}
e^{ip^+xy_1^-}
\int {d y^-_2 d y^-_3 \over 2\pi}
\theta(y^-_2-y_1^-)\theta(y_3^-) 
\nonumber \\
&\times & {1 \over 2} \langle p_A|{\bar q}(0)
\gamma^+ F^{\alpha +}(y_3^-)
{F^{+}}_\alpha(y^-_2) q(y_1^-)|p_A\rangle
\label{qkme}
\end{eqnarray}
where $|p_A\rangle$ is the relevant nuclear state.
The variable $x$ here is the fractional momentum associated with the
hard parton from the target that initiates the process.
Similar gluon-gluon correlation function $T_{g}(x,A)$ is also
important~\cite{LQS1}. 
In this type of the twist-4 parton-parton correlation functions, two 
integrals over the $y^-$ and $y_2^-$ can grow with the nuclear radius.  
However, if we require local color confinement, the
difference between the light-cone coordinates of the two field
strengths should be limited to the nucleon size and only one
of the two $y_i^-$ integrals can be extended to the size of nuclear
target.  The twist-4 parton-parton correlation functions are then
proportional to the size of the target, that is, 
enhanced by $A^{1/3}$.

It is important to emphasize that
using a pole in the complex $x_i$ (longitudinal momentum)
space to do the integral does not 
assume on-shell propagation for the scattered quark.  
Indeed, the $x_i$ integrals are not pinched
between coalescing singularities at such points, and the same
results could be derived by performing the $x_i$ integrals
without going through the $x_i=0$ points \cite{Qiu:2001hj}.

\section{Coherent power corrections in DIS}
\label{sec:3}

\begin{figure}
\begin{center}
  \epsfig{figure=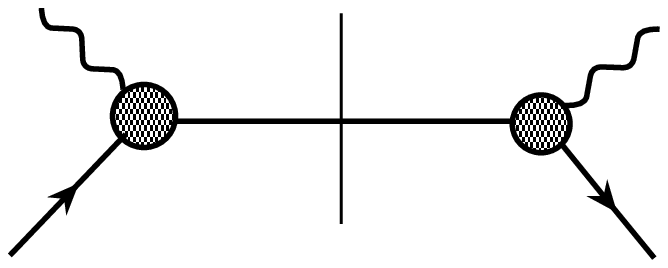,width=0.11\textwidth}
  \hskip 0.2in  
  \epsfig{figure=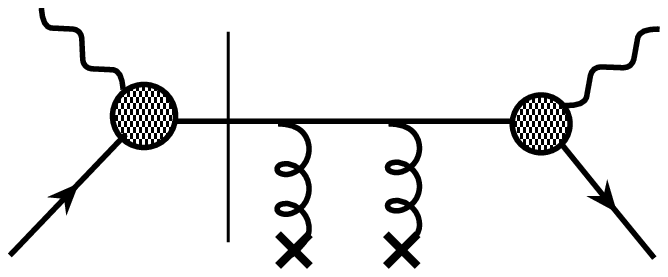,width=0.12\textwidth}
  \hskip 0.1in
  \epsfig{figure=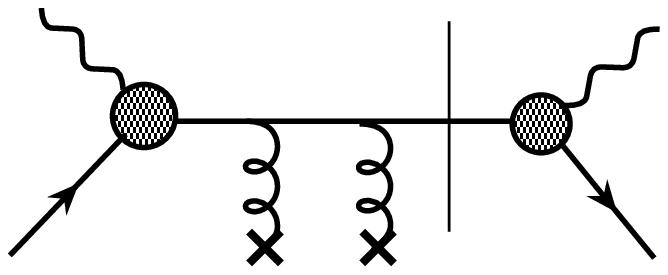,width=0.12\textwidth}

  (a) \hskip 1.7in (b)
 
  \epsfig{figure=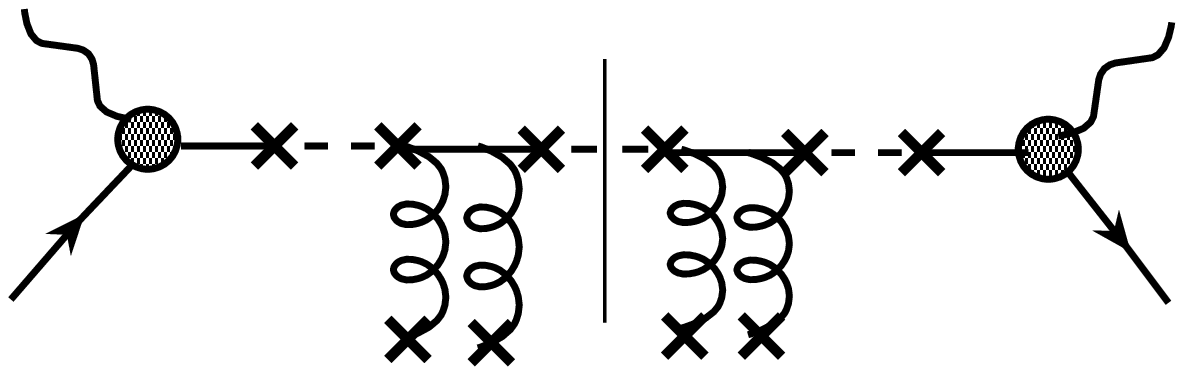,width=0.22\textwidth}
  
  (c)
 \caption{Tree diagrams give to the leading medium-size enhanced
          contributions to DIS cross sections.}
\label{fig4}
\end{center}
\end{figure}


The inclusive DIS cross section on a nucleus provides an unique probe
for effects of {\it coherent} multiple scatterings by varying the
value of Bjorken $x_B$.  
Under the approximation of one-photon exchange, the unpolarized
inclusive DIS cross section probes two structure functions: 
$F_T(x_B,Q^2)$ and $F_L(x_B,Q^2)$, corresponding to the transverse
and longitudinal polarization states of the virtual photon,
respectively \cite{Brock:1993sz,Guo:2001tz}. 
The structure functions at the lowest order in $\alpha_s$ are given
by~\cite{Brock:1993sz} 
\begin{eqnarray}
\label{FTLT}
F_T^{\rm (LT)}(x_B,Q^2) 
&=& \frac{1}{2} 
\sum_q e_q^2\, q(x_B,Q^2) 
        + {\cal O}(\alpha_s) \; ,
\\
F_L^{\rm (LT)}(x_B,Q^2) 
&=& {\cal O}(\alpha_s) \;,  
\end{eqnarray}
where (LT) indicates the leading twist contribution -- the first row
in Eq.~(\ref{facdis}), $\sum_q$ runs over the (anti)quark flavors,  
$e_q$ is their fractional charge, and $q(x,Q^2)$ is the leading twist
quark distribution:  
\begin{equation}
q(x,Q^2) =
\int \frac{d y^- }{2\pi}\, e^{i x p^+ y^-}
\langle p |\bar{q}(0)\, \frac{\gamma^+}{2}\, q(y^-) | p \rangle
\label{qDF}
\end{equation}
in the lightcone $A^+ = n^{\mu} A_{\mu}=0$ gauge for hadron momentum 
$p^\mu = p^+ \bar{n}^\mu$ with $\bar{n}^\mu = [1,0,0_\perp]$ and  
$n^\mu = [0,1,0_\perp]$.

The Feynman diagrams in Fig.~\ref{fig4} give the leading tree level
contributions to the lepton-nucleus DIS cross section.
The cut-line represents the final state \cite{Qiu:2004id}. 
For transversely polarized photons Fig.~\ref{fig4}(a) gives the
leading twist partonic contribution, 
\beq
d\hat{\sigma}^{(0)}_T
=\frac{1}{2}\, e_q^2 \, \delta(x-x_B)\, ,
\label{sigmaT0}
\eeq
from which $F_T^{\rm (LT)}$ in Eq.~(\ref{FTLT}) is derived after
convoluting with the leading twist quark distribution in
Eq.~(\ref{qDF}).  Diagrams with two gluons in Fig.~\ref{fig4}(b) 
gives the first power correction to transverse partonic cross section 
\cite{Qiu:2004id,Guo:2001tz}, 
\beq
d\hat{\sigma}^{(1)}_T
= \frac{1}{2}\, e_q^2 \left[\frac{1}{2N_c}\right]
  \frac{g^2}{Q^2} \left[2\pi^2 \tilde{F}^2(0)\right]
   x_B\left[- \frac{d}{dx}\delta(x-x_B)\right]
\label{sigmaT1}
\eeq
with the two-gluon field operator 
\beq
\left[\tilde{F}^2(0)\right]
=\int \frac{dy_2^- dy_1^-}{(2\pi)^2}
\left[F^{+\alpha}(y_2^-) F^{\ +}_{\alpha}(y_1^-)\right]
\theta(y_2^-) \, ;
\label{twogluon}
\eeq
and the first power correction to transverse structure function
\cite{Qiu:2004id,Guo:2001tz}, 
\beqa
F_T^{(1)}(x_B,Q^2)
&=&
\left[ \frac{4\pi^2 \alpha_s}{Q^2}
       \left(\frac{1}{2N_c}\right) \right] 
\nonumber\\
&\times & 
\frac{1}{2}\,
\sum_q\, e_q^2\, x_B\, \frac{d}{dx_B}T_q(x_B,A) 
\label{delFT1}
\eeqa
with correlation function, $T_{q}(x_B,A)$, given in
Eq.~(\ref{qkme}).
  
The four-parton correlation functions $T_q$'s
are nonperturbative and must be taken from experiments.  
To estimate the magnitude of $T_q$'s, 
we could choose a simple ansatz \cite{LQS} 
\beq
T_i(x,A)=\lambda^2\, A^{1/3}\, \phi_{i/p_A}(x,A)
\label{Tqans}
\eeq
for $i=q,g$ in terms of the corresponding twist-2 effective nuclear
parton distribution $\phi_{i/A}$.
We choose this form because we expect the $x$-dependence of
the probability to detect the hard parton to
be essentially unaffected by the presence or absence of an
additional soft scattering.    
In Eq.~(\ref{Tqans}) $\lambda$ is assumed to be a constant with
dimensions of mass.  This ansatz facilitates the comparison to
data \cite{Qiu:2001hj}.   

\begin{figure}
\begin{center}
  \epsfig{figure=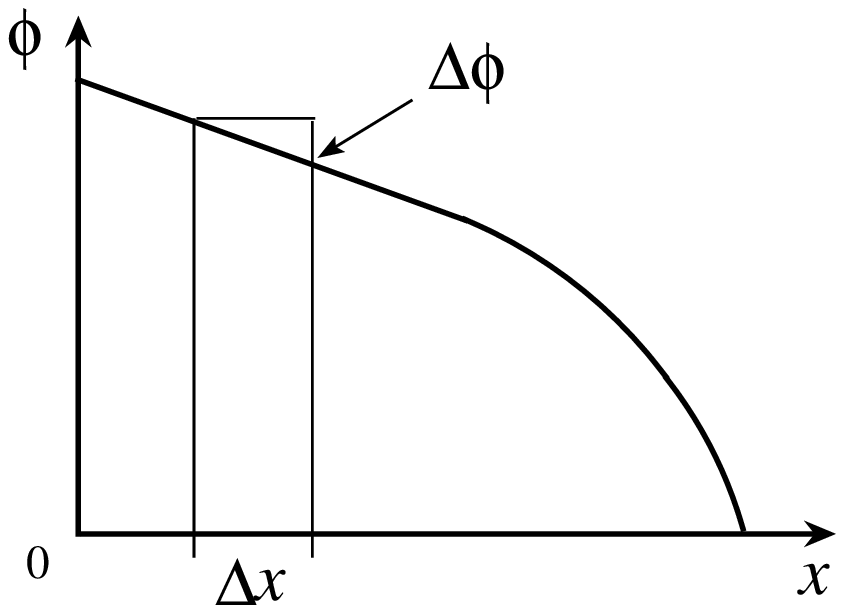,width=0.21\textwidth}
   \hskip 0.06in
  \epsfig{figure=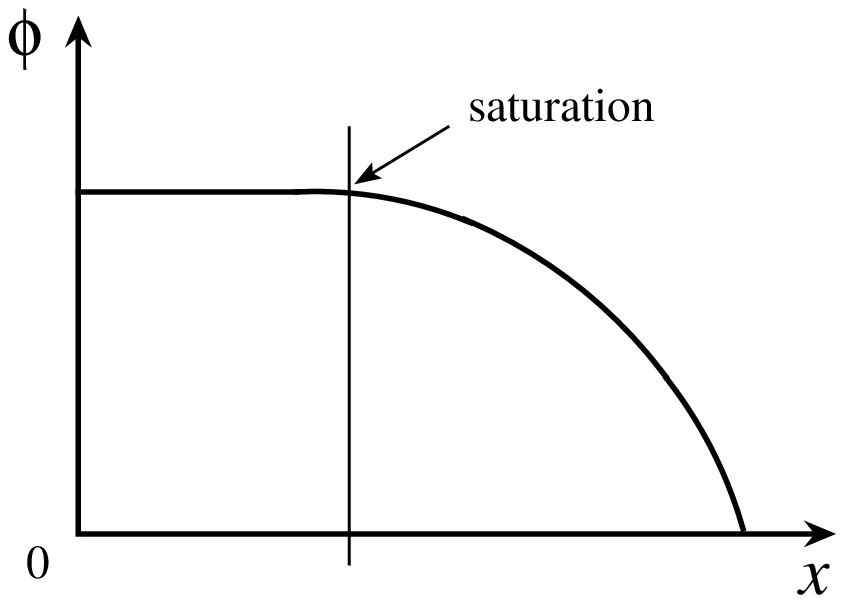,width=0.21\textwidth}

  (a) \hskip 1.4in (b)
\caption{Sketch for the $x$-dependence of normal (a) and saturated (b)
  parton distributions. }
\label{fig5}
\end{center}
\end{figure}


If we further assume that a nucleus is made of a group of loosely
bound color singlet nucleons, packed in a hard sphere of radius $R
A^{1/3}$, we can approximate the matrix element of nuclear state in 
Eq.~(\ref{qkme}) into a product of matrix elements of nucleon 
states \cite{Qiu:2004id,AM},
\beqa
&& \langle p_A | \bar{q}(0) \frac{\gamma^+}{2}
             \left[\tilde{F}^2(0)\right] q(y^-) |p_A\rangle
\\
&& \approx 
   \frac{A}{2\pi} 
   \left[ \frac{\frac{3}{4}R A^{1/3}}
               {\frac{1}{\frac{4\pi}{3}R^3}}\right]
   \langle F^{+\alpha}F^{\ +}_{\alpha} \rangle\,
   \langle p| \bar{q}(0) \frac{\gamma^+}{2} q(y^-) |p\rangle 
\nonumber
\label{matrix-4}
\eeqa
with $p=p_A/A$ and the two-gluon matrix element
\cite{Qiu:2004id} 
\beq
\langle F^{+\alpha}F^{\ +}_{\alpha} \rangle
\equiv
\frac{1}{p^+} \int \frac{dy^-}{2\pi}
\langle p|F^{+\alpha}(0)F^{\ +}_{\alpha}(y^-)|p \rangle
\theta(y^-)\, .
\label{htscale}
\eeq
Substituting Eq.~(\ref{matrix-4}) into the definition of
quark-gluon correlation function, $T_q(x,A)$ in Eq.~(\ref{qkme}), we
drive 
\beq
\lambda^2 \approx \left(\frac{3}{4}R\right)
            \left(\frac{1}{\frac{4\pi}{3}R^3}\right)
            \langle F^{+\alpha}F^{\ +}_{\alpha} \rangle \, ,
\label{lambda2}
\eeq
and
\beqa
F_T^{(1)}(x_B,Q^2)
& \approx &
\frac{3\pi\alpha_s}{8Q^2R^2} 
\langle F^{+\alpha}F^{\ +}_{\alpha} \rangle
\left(A^{1/3}-1\right)
\nonumber\\
&\times & 
\frac{1}{2}\,
\sum_q\, e_q^2\, x_B\, \frac{d}{dx_B} q_A(x_B,Q^2) 
\label{FT1}
\eeqa
where $q_A$ defined in Eq.~(\ref{qDF}) with $p=p_A/A$.
In deriving Eq.~(\ref{FT1}), we used $A^{1/3}-1$ instead of $A^{1/3}$
for the line integral between the two pairs of field operators in
Eq.~(\ref{matrix-4}) so that the nuclear effect vanishes for $A=1$
\cite{Qiu:2004id}. 
From Eq.~(\ref{FT1}) and known generic $x$-dependence of parton
distributions, as sketched in Fig.~\ref{fig5}(a), 
the leading coherent power correction {\it suppresses} the DIS cross
section or structure functions at small $x_B$.

If $x_B$ is small enough, the virtual photon could interact
{\it coherently} with all nucleons inside a large nucleus.  
From the Feynman diagrams in Fig.~\ref{fig4}(c) with
$2N$ gluons, we derive the leading Nth coherent power corrections  
to the transverse structure function \cite{Qiu:2004id}, 
\beqa
F_T^{(N)}(x_B,Q^2)
&=&
\left[\frac{3\pi\alpha_s}{8Q^2R^2} 
\langle F^{+\alpha}F^{\ +}_{\alpha} \rangle
\left(A^{1/3}-1\right)\right]^N
\nonumber\\
&\times & 
\frac{1}{2}\,
\sum_q\, e_q^2\, \frac{1}{N!}\, 
x_B^N\, \frac{d^N}{dx_B^N} q_A(x_B,Q^2) \, .
\label{FTN}
\eeqa
Summing over all leading $A^{1/3}$-enhanced power corrections, the
first column of the right-hand-side of Eq.~(\ref{facdis}), we obtain
\cite{Qiu:2004id} 
\beqa
F_T^A(x_B,Q^2)
&=&
\sum_{n=0}^{N} \frac{1}{n!}\
\left[\frac{\xi^2}{Q^2}\left(A^{1/3}-1\right)\right]^n
\nonumber \\
&& \times \
x_B^n \frac{d^n}{dx_B^n}F_T^{\rm A(LT)}(x_B,Q^2)
\label{FTAN}
\eeqa
with a characteristic scale of the power corrections
\beqa
\xi^2
&\equiv & 
\frac{3\pi \alpha_s}{8R^2} \,
\langle F^{+\alpha}F^{+}_{\ \alpha}\rangle\, .
\label{xi2}
\eeqa
If we approximate $N\sim \infty$ in Eq.~(\ref{FTAN}), we obtain,
\beq
F_T^A(x_B,Q^2)
\approx 
F_T^{\rm A(LT)}(x_B(1+\Delta),Q^2)
\label{FTA}
\eeq
with $\Delta \equiv  (A^{1/3}-1)\,\xi^2/Q^2$, a shift in $x_B$.
Similar expression was derived for the longitudinal structure function
$F_L^A(x_B,Q^2)$ \cite{Qiu:2004id}.  


\begin{figure}
\begin{center}
  \epsfig{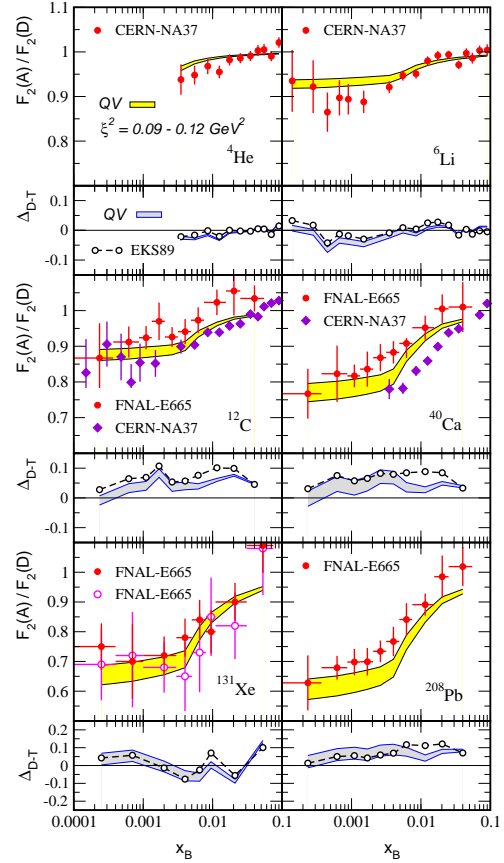}
\caption{From~\cite{Qiu:2004id}, all-twist resummed $F_2(A)/F_2(D)$
  calculation versus DIS data on nuclei~\cite{DISdata}. 
Data-Theory $\Delta_{D-T}$ is also shown. }
\label{fig6}
\end{center}
\end{figure}


With only one unknown matrix element, 
$\langle F^{+\alpha}F^{\ +}_{\alpha} \rangle$,
our calculated results can be
easily tested and challenged.  From Eq.~(\ref{FTA}), the
nuclear dependence in structure functions should come from two
distinctive sources: universal $A$-dependence in leading twist parton
distributions and process sensitive $A$-dependence from power
corrections \cite{Qiu:2001hj,Qiu:2004id}.  By comparing our numerical
results, evaluated with CTEQ6 PDFs~\cite{Pumplin:2002vw}, with
the data, we extract the {\it maximum} size of power corrections.
For $\xi^2=0.09-0.12$~GeV$^2$, our calculated reduction in nuclear
structure functions is consistent with the $x_B$-, $Q^2$- and
$A$-dependence of the data as demonstrated Fig.~\ref{fig6}. 

The predictive power of factorization approach resides in the
universality of unknown matrix elements.  
Without additional unknown matrix elements, our predictions for
neutrino DIS cross section on an iron target are 
consistent with NuTeV data \cite{Qiu:2004qk,vitev-dis}.  
At small and moderate $Q^2$, our prediction gives a very good
description of SLAC E143 data on $R=\sigma_L/\sigma_T$ of 
$^{12}$C target~\cite{Qiu:2004id,Abe:1998ym}

The resummed power correction has a nontrivial dependence on the
target $A$ and kinematic variables $x_B$ and $Q^2$ because its
connection to the $x$-dependence of the active parton distribution.
The effect disappears if the parton distribution is  
saturated at a very small $x$, as sketched in Fig.~\ref{fig5}(b).  
The leading power correction calculated here is a consequence of 
the limited phase space in a given collision.  If we would
integrate over $x_B$ from $-\infty$ (or zero) to $\infty$ for a purely
inclusive process (or a process with an infinite collision energy),
the $F_T^{(N)}$ vanishes for $N\neq 0$ due to the unitarity
condition \cite{Qiu:1990xy}.  

\begin{figure}
\begin{center}
  \epsfig{figure=dis_single.ps,width=0.12\textwidth}
  \hskip 0.2in
  \epsfig{figure=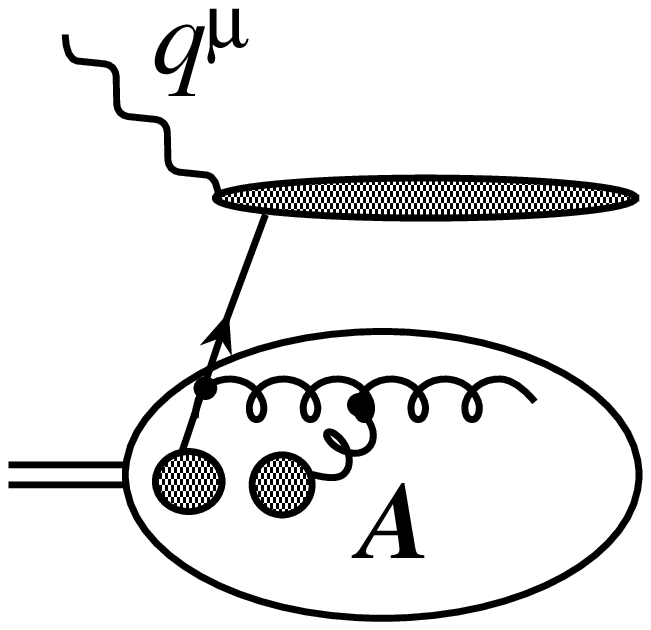,width=0.10\textwidth}
  \hskip 0.2in
  \epsfig{figure=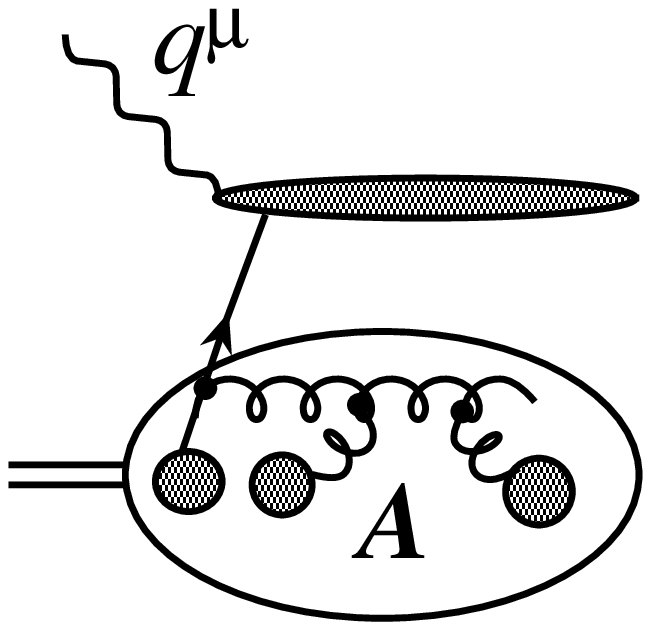,width=0.10\textwidth}

  (a) \hskip 0.9in (b) \hskip 0.7in (c)
\caption{Sample DIS diagrams have initial-state collinear divergence
  involving parton recombinations. }
\label{fig7}
\end{center}
\end{figure}

QCD factorization systematically moves all collinear divergences
from partonic scattering into long-distance matrix
elements. At the leading power, DGLAP evolved parton distributions are
sufficient to absorb all collinear divergences in $\gamma^*$-parton
scattering cross section to ensure the infra-safety of all coefficient
functions in the first row of Eq.~(\ref{facdis}).  However, DGLAP
evolved parton distributions do not remove collinear divergences
involving multiple parton recombination, like those in
Fig.~\ref{fig7}(b) and (c).  A modified evolution equation taking into
account the parton recombination slows down the fast growth of parton
density at small $x$ \cite{AM,GLR}, and keeps a positive density of 
small-$x$ gluons at low $Q^2$ \cite{Eskola-NPB}.  All order 
resummed power corrections -- coherent parton recombinations to parton
evolution should be vary valuable for approaching the region of parton
saturation from the perturbative side and for getting a reliable
picture of small-$x$ partons \cite{Zhu:2004zw,Qiu-Kang}.


\section{Power correction to Drell-Yan cross section}
\label{sec:4}

\begin{figure}
\begin{center}
\begin{minipage}[c]{1.4in}
\begin{center}
  \epsfig{figure=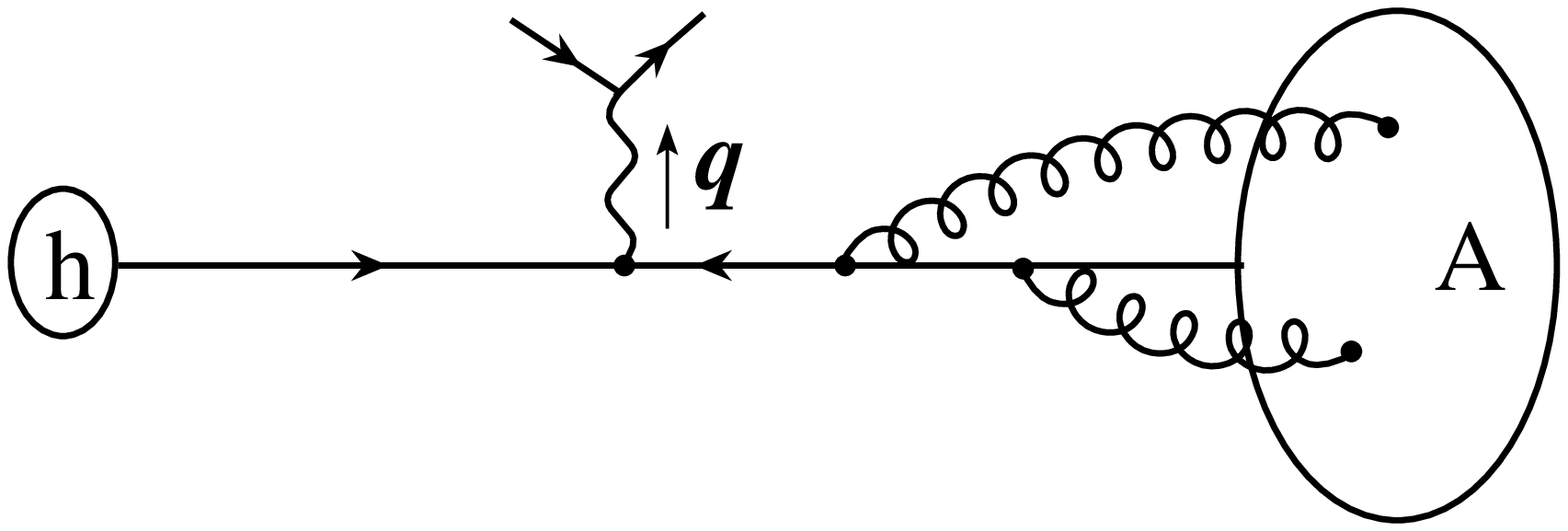,width=1.4in}\\
  {\scriptsize (a)}\\
  \vskip 0.1in
  \epsfig{figure=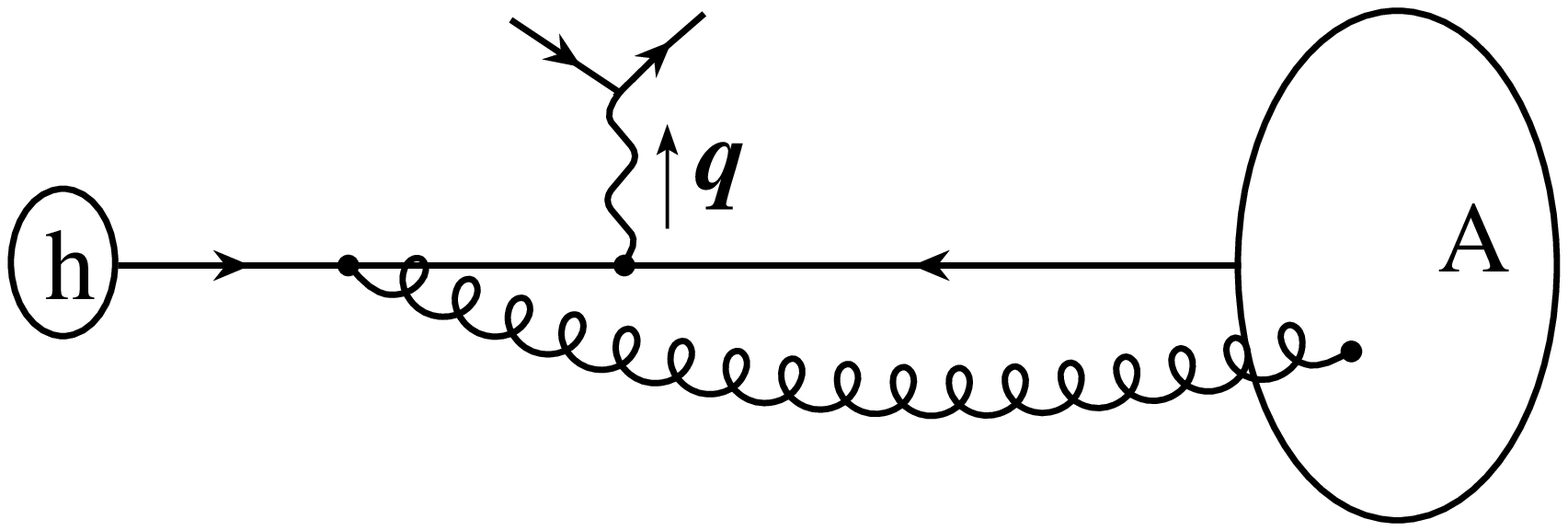,width=1.4in}\\
  {\scriptsize (b)}
\end{center}
\end{minipage}
\hfil
\begin{minipage}[c]{1.8in}
\begin{center}
  \epsfig{figure=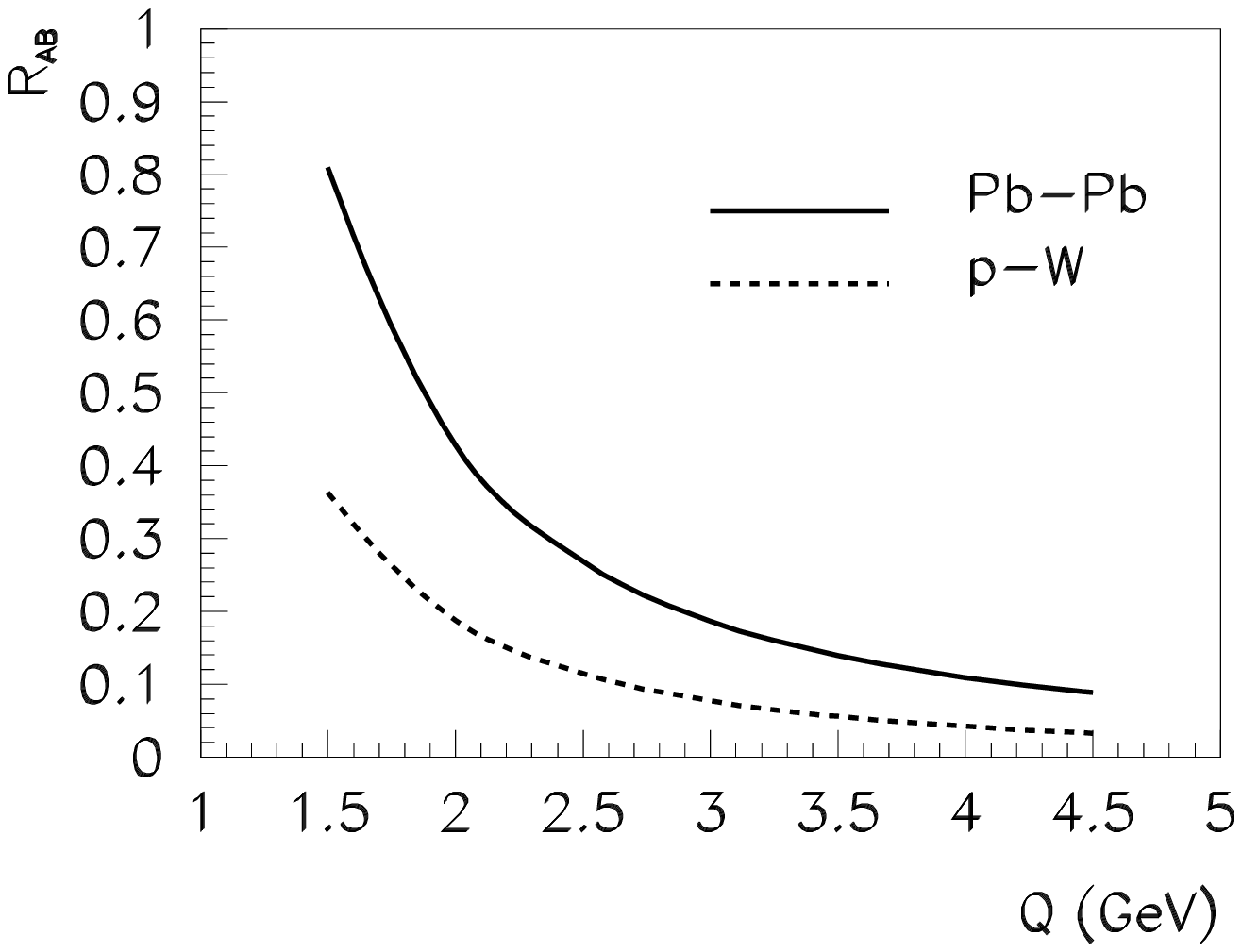,width=1.8in}\\
  {\scriptsize (c)}
\end{center}
\end{minipage}
\caption{Sketch for Drell-Yan process in hadron-nucleus collision with
multiple interactions internal to nucleus (a) and sensitive to hard
scattering (b); and the nuclear modification factor, $R_{AB}(Q)$, as a
function of dilepton mass $Q$.} 
\label{fig8}
\end{center}
\end{figure}


It was shown by the NA38 and NA50 Collaborations that muon
pair production for dimuon invariant mass between $\phi$ and
J/$\psi$, known as the intermediate mass region (IMR), in heavy
nucleus-nucleus collisions exceeds the expectation 
based on a linear extrapolation of the $p-A$ sources with the product
of the mass numbers of the projectile and target nuclei
\cite{NA38NA50-dimuon}.  The excess increases with the number of
participant nucleons, and the ratio between the observed dimuon yield
and the expected sources reaches a factor of 2 for central Pb-Pb
collisions.  There have been a lot of effort to attribute such an
excess to the enhancement of open charm production \cite{OpenCharm},
thermal dimuons production \cite{thermal-l}, and secondary
meson-meson scattering in nuclear medium \cite{Meson-meson}.  As shown
in Ref.~\cite{NA38NA50-dimuon}, the Drell-Yan continuum is the
dominant source of the dilepton production in this mass
range.  An enhancement in the Drell-Yan continuum is much more
effective than all other sources for interpreting the observed excess.

In hadron-nucleus and nucleus-nucleus collisions, more partons are
available at a given impact parameter.  
Before the hard collision of producing
the lepton-pair, partons from different nucleons can either interact
between themselves, as sketched in Fig.~\ref{fig8}(a) which leads to
the universal nuclear dependence in parton distributions, or interact
with the incoming parton, as sketched in Fig.~\ref{fig8}(b) which 
gives the medium-size enhanced power corrections \cite{Qiu:2001zj}.
For the kinematics of IMR Drell-Yan process, parton distributions have
a week nuclear dependence and produce a small reduction to the
cross section.  On the other hand, the medium size enhanced power
corrections are process dependent, and actually {\it increase} the
production rate and become more important when $Q^2$ decreases
\cite{Qiu:2001zj}.  Let inclusive Drell-Yan cross section in nuclear
collisions be approximated as
\begin{eqnarray}
\frac{d\sigma_{AB}}{dQ^2} 
&\approx &
AB\frac{d\sigma_{NN}^{(S)}}{dQ^2}
+ \frac{d\sigma_{AB}^{(D)}}{dQ^2} + \dots
\nonumber \\
& \equiv &
AB\frac{d\sigma_{NN}^{(S)}}{dQ^2}
\left[ 1 + R_{AB}(Q) \right]\, ,
\label{dy-s-d}
\end{eqnarray}
where superscripts $(S)$ and $(D)$ represent the single and double
scattering, respectively.  The $R_{AB}(Q)$ defines a nuclear
modification factor of Drell-Yan continuum.  
Fig.~\ref{fig8}(c) shows the calculated $R_{AB}(Q)$ from medium-size
enhanced double scattering \cite{Qiu:2001zj}.  In evaluating
Fig.~\ref{fig8}(c), we used the same quark-gluon correlation function,
$T_{q}(x,A)$, and neglected the $A$-dependence in parton
distributions.  Therefore, we expect that the true enhancement to the
Drell-Yan continuum might be slightly smaller than $R_{AB}(Q)$ in
Fig.~\ref{fig8}(c), which is a very significant effect.

\begin{figure}
\begin{center}
  \epsfig{figure=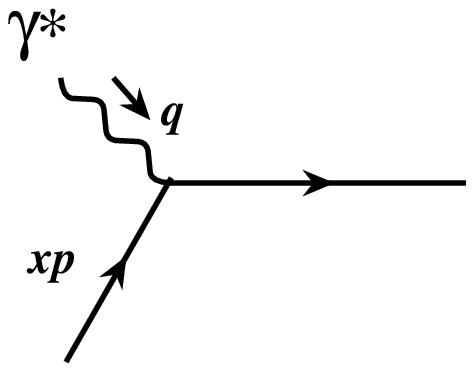,width=0.1\textwidth}
  \hskip 0.2in
  \epsfig{figure=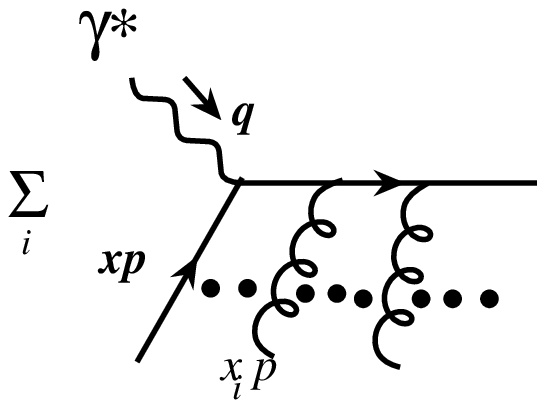,width=0.11\textwidth}
  \hskip 0.2in
  \epsfig{figure=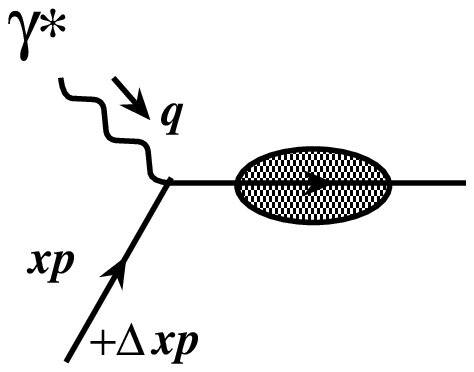,width=0.10\textwidth}
  
  (a) \hskip 0.8in (b) \hskip 0.8in (c)
\caption{DIS with a space-like hard scale}
\label{fig9}
\end{center}
\end{figure}

\begin{figure}
\begin{center}
  \epsfig{figure=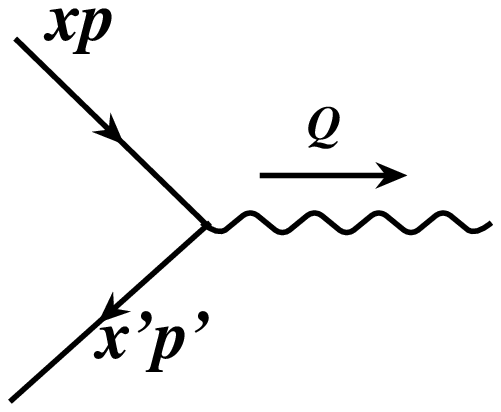,width=0.09\textwidth}
  \hskip 0.2in
  \epsfig{figure=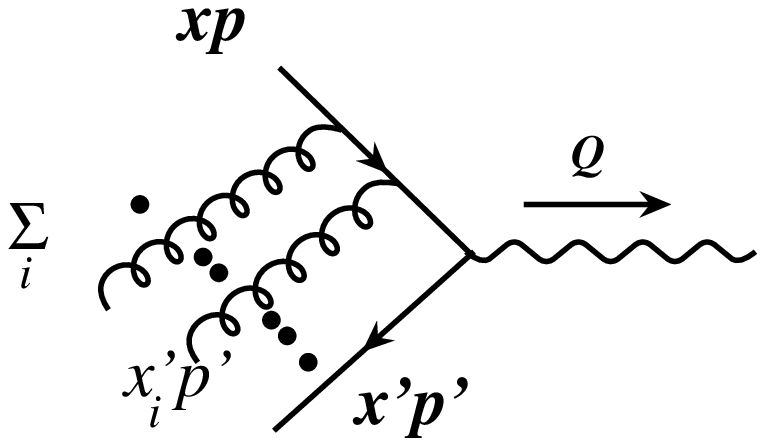,width=0.13\textwidth}
  \hskip 0.2in
  \epsfig{figure=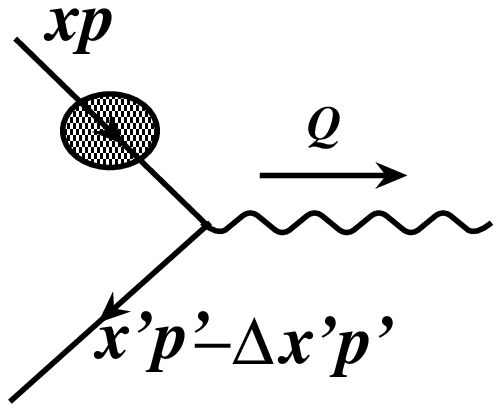,width=0.10\textwidth}

  (a) \hskip 0.8in (b) \hskip 0.8in (c)
\caption{Drell-Yan with a time-like hard scale}
\label{fig10}
\end{center}
\end{figure}


It is perhaps surprising that the medium-size enhanced power
correction to DIS and Drell-Yan cross sections carries a different
sign.  The sign difference is a consequence of the kinematic nature of 
the calculated leading power corrections.  The sketch in
Fig.~\ref{fig9} shows that the sum of multiple soft rescattering to
the final-state parton in DIS effectively broadens the parton's
transverse momentum \cite{Guo:1998rd} (or gives the parton an
effective mass \cite{Qiu:2004qk}); and the on-shell condition of the
final-state parton forces an extra momentum fraction $\Delta xp$ to
the incoming parton.  On the other hand, the multiple initial-state
soft rescattering in Drell-Yan process {\it broadens} the incoming
parton's momentum, and the kinematics (or the measured $q^\mu$) of the
observed dilepton pair forces the active parton from the target to
take away some of its momentum, a negative shift in momentum fraction,
as shown in Fig.~\ref{fig10}(c).


\section{Transverse momentum $Q_T$ broadening}
\label{sec:5}

Coherent multiple rescatterings not only modify the production rate of 
inclusive cross sections, but also affect the momentum distribution of
produced particles.  Although the amount of broadening due to each
soft rescattering is too weak a scale to warrant a reliable
calculation, an averaged broadening in a hard collision could be a
physical quantity calculable in pQCD \cite{Qiu:2001hj}.  
 
\begin{figure}
\begin{center}
  \epsfig{figure=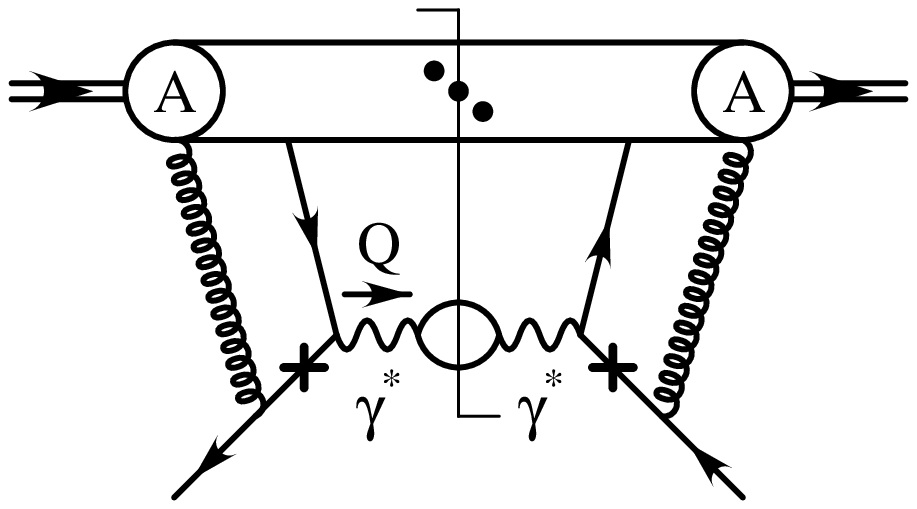,width=1.3in}
  \hskip 0.3in
  \epsfig{figure=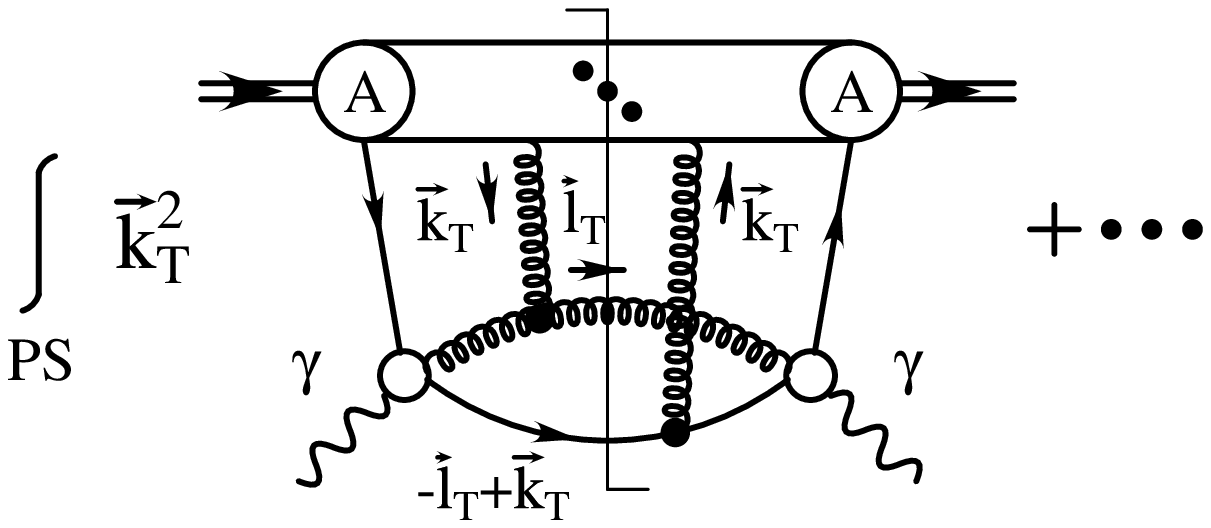,width=1.65in}\\
  {\scriptsize (a) \hskip 1.6in (b)}
\end{center}
\caption{(a) Lowest order diagram that contributes to the Drell-Yan
transverse momentum broadening in hadron-nucleus collisions;
(b) Sketch for the lowest order contributions to the averaged
di-jet momentum imbalance in photoproduction.}
\label{fig11}
\end{figure}

In Ref.~\cite{Guo:1998rd}, Drell-Yan transverse momentum broadening
was calculated in pQCD by evaluating the lowest order soft
rescattering diagram in Fig.~\ref{fig11}(a).  Although the direct $Q_T$
modification from soft rescattering to $d\sigma/dQ^2 dQ_T^2$ is not
perturbatively calculable because of the size of small $Q_T$ kick, the
$Q_T$ broadening,
\beq 
\langle Q_T^2 \rangle
\equiv 
\left. \int dQ_T^2\, Q_T^2\, \frac{d\sigma}{dQ^2 dQ_T^2}
\right/ \int dQ_T^2\, \frac{d\sigma}{dQ^2 dQ_T^2}\, ,
\label{QT2}
\eeq
is calculable
\cite{Qiu:2001hj,Qiu:1990xy}.  It was found that the Drell-Yan
transverse momentum broadening in hadron-nucleus collisions can be 
expressed in terms of the same quark-gluon correlation
function $T_q(x,A)$ (not its derivative) \cite{Guo:1998rd},
\beq
\langle Q_T^2 \rangle
\approx 
\left(\frac{4\pi^2 \alpha_s}{3}\right)\,
\frac{\sum_q e_q^2 \int dx' f_{\bar{q}/h}(x')\, T_q(\tau/x',A)/x'}
      {\sum_q e_q^2 \int dx' f_{\bar{q}/h}(x')\, f_q(\tau/x',A)/x'}
\label{dy-qt2}
\eeq
where $\sum_q$ runs over all quark and antiquark flavors, $e_q$ is
the quark fractional charge, and $\tau=Q^2/s$, in terms of the
lepton-pair invariant mass $Q$ and hadron-hadron center of mass
energy, $\sqrt{s}$. 

Adopting the model in Eq.~(\ref{Tqans}), the lowest order Drell-Yan
transverse momentum broadening in Eq.~(\ref{dy-qt2}) can be simplified
as
\beq
\langle Q_T^2 \rangle_{4/3}
= \left(\frac{4\pi^2 \alpha_s}{3}\right)\,
   \lambda^2\,A^{1/3}\, .
\label{dy-qt2-lqs}
\eeq
By comparing Eq.~(\ref{dy-qt2-lqs}) to data from Fermilab E772 and
CERN NA10 experiments \cite{E772-dy,NA10-dy}, it was found
\cite{Guo:1998rd} that $\lambda_{\rm DY}^2 \sim 0.01$~GeV$^2$, which
corresponds to $\xi^2_{\rm DY} \sim 0.04$~GeV$^2$ that is about a
factor of 2 {\it smaller} from what was extracted from inclusive DIS
data, where the leading twist nuclear dependence was not included.
The difference might be caused by the difference between initial-state
and final-state rescattering effects \cite{Qiu:2001hj}. 

Momentum imbalance between two final-state jets in photoproduction off 
a nuclear target was calculated by evaluating the diagrams like one in
Fig.~\ref{fig11}(b), and was compared with Fermilab E683 data
\cite{LQS}, with an assumption that the momentum imbalance between two
jets is approximately the same as the momentum imbalance between two
final-state partons.  Again, the calculated momentum imbalance was
expressed in terms of the same quark-gluon and gluon-gluon correlation
function \cite{LQS}.  By comparing with the momentum imbalance data
for jet transverse momentum $p_T>4$~GeV \cite{E683}, it was found
\cite{LQS} that $\lambda^2_{\rm dijet}=0.05-0.1$~GeV$^2$, which
corresponds to $\xi^2\sim 0.2$~GeV$^2$ that is about a factor 2 {\it
  bigger} than what was found in DIS.  As pointed out in
Ref.~\cite{Qiu:2001hj}, high order corrections to initial-state and
final-state soft rescattering could be very different and
significantly reduce the difference of the nonperturbative parameter.

\section{Nuclear suppression in hadron-nucleus collisions}
\label{sec:6}

Dynamical nuclear effects induced by soft rescattering can be studied
through the ratio of particle production rates in hadron-nucleus
and hadron-hadron collisions.  However, there is no observed variable,
like $x_B$ in inclusive DIS, that can directly measure the coherence
length of the hard scattering.  All incoming parton momentum fractions
are convoluted over, and strength of the collision is measured by
Lorentz invariants, like Mandelstam variables, $\hat{s}$, $\hat{t}$,
and $\hat{u}$.  Unlike in DIS and in Drell-Yan, there are both
initial-state and final-state soft rescattering in hadronic
collisions, which could in principle lead to medium size enhanced
power corrections.  Since the hard scales, $\hat{s}$, $\hat{t}$,
and $\hat{u}$, in hadronic collisions are often much larger than a
couple of GeV$^2$, effect of the medium size enhanced power
corrections to hard probes in hardonic collisions is in general less
significant.  However, in the most forward (backward) region, the
invariant $\hat{t}$ ($\hat{u}$) could be much smaller than the other
invariants, so that the power corrections in $1/\hat{t}$ ($1/\hat{u}$)
could become very important \cite{Qiu:2004da}.  

Consider, for example, the
single hadron inclusive production in hadron-nucleus collision
as shown in Fig.~\ref{fig2}(b).  Once we fix the momentum fractions
$x_a$ and $z_1$, the effective interaction region is determined by the
momentum exchange $q^{\mu}=(x_aP_a-P_c/z_1)^{\mu}$.
In the head-on frame of $q-P_b$, the scattered parton of momentum
$\ell$ interacts coherently with partons from different nucleons at
the same impact parameter, just like that in DIS.
Interactions that take place between the partons from the 
nucleus and the incoming parton of momentum $x_aP_a$ and/or the 
outgoing parton of momentum $P_c/z_1$ at a different impact parameter
are much less coherent and actually dominated by the independent
elastic scattering \cite{Gyulassy:2002yv}. 
Similar to the DIS case \cite{Qiu:2003vd}, we find \cite{Qiu:2004da}
that resumming the coherent scattering with multiple nucleons is
equivalent to a shift of the momentum fraction of the active parton
from the nucleus in Fig.~\ref{fig2}(b),
\begin{equation}
x_b\rightarrow x_b 
\left (1+ \frac{C_d \xi^2(A^{1/3}-1)}{(-\hat{t})} \right) \, ,
\label{shift-pa}
\end{equation}
with the hard scale $\hat{t}=q^2=(x_aP_a-P_c/z_1)^2$ and the color factor
$C_d$ depending on the flavor of parton $d$. $C_{q(\bar{q})}=1$ and 
$C_g=C_A/C_F=9/4$ for quark (antiquark) and gluon, respectively.
The shift in Eq.~(\ref{shift-pa}) leads to a net suppression of the
cross sections, and the $\hat{t}$-dependence of this shift indicates
that the attenuation increases in the forward rapidity region.  

\begin{figure}
\begin{center}
  \epsfig{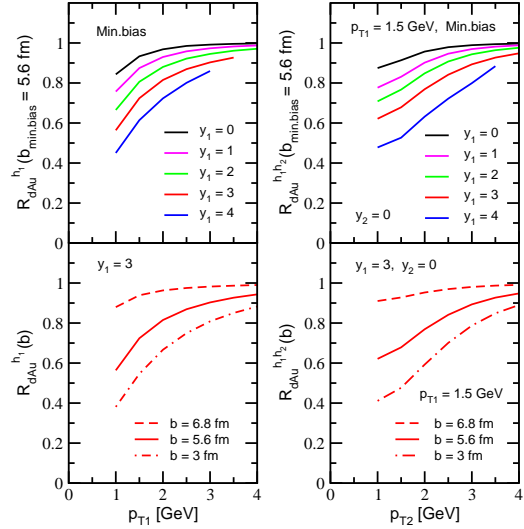}
\caption{From \cite{Qiu:2004da}. Suppression of the single
  and double inclusive hadron production rates as a function of $p_T$
  in $d+Au$ collisions at  $\sqrt{s}=200$GeV$^2$ at different
  rapidities (Top) and at different impact parameters (Bottom).  The
  trigger hadron $p_{T_1}=1.5$~GeV, $y=3$, and the associated hadron
  at $y=0$.}
\label{fig12}
\end{center}
\end{figure}

The top panels of Fig.~\ref{fig12} show the rapidity and
transverse momentum dependence of $R^{h_1}_{dAu}(b)$ and 
$R^{h_1h_2}_{dAu}(b)$, the ratio of single and double
hadron production, respectively, in minimum bias $d+Au$ collisions. 
The amplification of the suppression effect at forward $y_1$ comes 
from  the steepening of the parton distribution 
functions at small $x_b$ and the decrease of the Mandelstam variable 
$(-\hat{t})$. At high $p_{T1},p_{T2} $ the attenuation is found to  
disappear in accord with the QCD factorization 
theorems \cite{CSS-fac}.  The bottom panels of Fig.~\ref{fig12} 
show the growth of the  nuclear attenuation effect with centrality.

Dihadron~correlations 
\ $C_2(\Delta \varphi) = 
\frac{1}{N_{\rm trig}} \frac{dN^{h_1h_2}_{\rm dijet}}{d\Delta \varphi} $ 
associated with $2 \rightarrow 2$
partonic hard scattering processes, after subtracting the bulk
many-body collision background, can be approximated by near-side
and away-side Gaussians. The acoplanarity, $\Delta \varphi  \neq \pi$,
arises from high order QCD corrections and in the presence of nuclear
matter - transverse momentum diffusion \cite{Gyulassy:2002yv}.  
If the strength of the away-side correlation function in elementary 
N+N collisions is normalized  to  unity, dynamical quark and gluon 
shadowing in $p+A$ reactions will be manifest in the  attenuation of  
the {\em area}  $A_{\rm Far} = R^{h_1h_2}_{pA}(b)$ \cite{Qiu:2004da}. 


\begin{figure}
\begin{center} 
  \epsfig{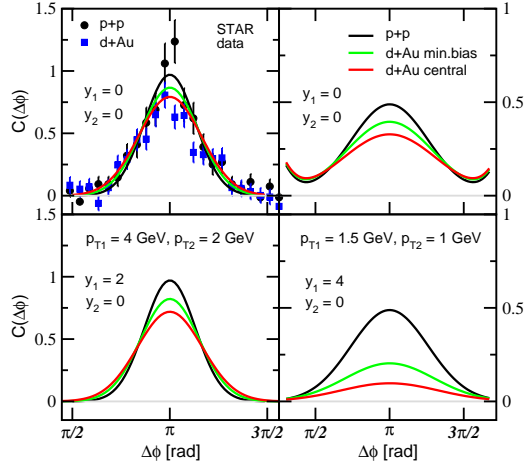}
\caption{From \protect\cite{Qiu:2004da}. Centrality dependence of 
$C_2(\Delta \varphi)$ at various rapidities and moderate (left) 
and small (right) transverse momenta.   
Central $d+Au$ and $p+p$ data  from STAR \protect\cite{Adams:2003im}.}  
\label{fig13}
\end{center} 
\end{figure}


The left panels of Fig.~\ref{fig13} 
show that for $p_{T_1}=4$~GeV, $p_{T_2}=2$~GeV 
the dominant effect in $C_2(\Delta \varphi)$ is a small increase 
of the broadening with centrality, compatible with the 
PHENIX\cite{Rak:2004gk} and STAR\cite{Adams:2003im} 
measurements. Even at forward rapidity, such as $y_1 = 2$, the 
effect of power corrections in this transverse momentum  range is  
not very significant. At small $p_{T_1}=1.5$~GeV,  $p_{T_2}=1$~GeV, 
shown in the right hand side of Fig.~\ref{fig13}, the apparent   
width of the away-side  $C_2(\Delta \varphi)$ is larger. 
In going from  midrapidity,  $y_1 = y_2 = 0$, to forward  rapidity, 
$y_1 = 4, y_2 = 0$,  we find a significant reduction by a factor 
of 3 - 4  in the strength of dihadron correlations. 
Preliminary STAR results \cite{Ogawa:2004sy} 
are consistent with our predictions.

\section{Summary and outlook}
\label{sec:7}

We have briefly reviewed a pQCD factorization
approach to coherent multiple scattering and argued that the
$A^{1/3}$-type medium size enhanced power corrections caused by
multiple soft rescattering can be consistently calculated in pQCD.
We presented calculations of rescattering effects in inclusive cross
sections as well as transverse momentum broadening (or the moment of
particle transverse momentum distributions.).
By studying the coherent multiple scattering, we can probe new sets of
fundamental and universal multiparton correlation functions in nuclear
medium.  These new functions provide new insights into the
nonperturbative regime of QCD.  

Initial pQCD calculations of multiple rescattering effects is very
successful in understanding nuclear dependence of inclusive cross
sections.  For a wide range of observables, the estimate for the only
unknown parameter is reasonably consistent.  With a
single unknown parameter, the calculations describe many observables
and their nuclear dependences fairly well.  
The factorization approach, with its intuitive and transparent results, 
can be easily applied to study the nuclear modification of other 
physical observables in $p+A$ reactions. The systematic incorporation 
of coherent power corrections provides a novel tool to address the 
most interesting transition region  between ``hard'' and ``soft'' 
physics in hadron-nucleus collisions.  


This work is supported in part by the U.S. Department of Energy under
Grant No. DE-FG02-87ER40371.



\begin{thebibliography}{}
%
%
\bibitem{YellowReport} 
CERN Yellow Report (CERN-2004-009 on ``Hard probes in heavy-ion
collisions at the LHC'', edited by M. Mangano, H. Satz, and
U. Wiedemann (Geneva, Switzerland, 2004), and references therein.

\bibitem{Qiu:2001hj}
  J.~W.~Qiu and G.~Sterman,
  Int.\ J.\ Mod.\ Phys.\ E {\bf 12} (2003) 149
  [arXiv:hep-ph/0111002].

\bibitem{Qiu:2004id}
  J.~W.~Qiu and I.~Vitev,
  arXiv:hep-ph/0410218.

\bibitem{BDMPS} 
  R.~Baier, Y.~L.~Dokshitzer, A.~H.~Mueller, S.~Peigne and D.~Schiff,
  Nucl.\ Phys.\ B {\bf 483} (1997) 291
  [arXiv:hep-ph/9607355];
  Nucl.\ Phys.\ B {\bf 484} (1997) 265
  [arXiv:hep-ph/9608322].

\bibitem{Zakharov} 
B.G. Zakharov, JETP Lett. 70 (1999) 176; 
  R.~Baier, D.~Schiff and B.~G.~Zakharov,
  Ann.\ Rev.\ Nucl.\ Part.\ Sci.\  {\bf 50} (2000) 37
  [arXiv:hep-ph/0002198].

\bibitem{Gyulassy}
  M.~Gyulassy, P.~Levai and I.~Vitev,
  Nucl.\ Phys.\ B {\bf 594} (2001) 371,
and references therein.

\bibitem{GW-loss}
  X.~F.~Guo and X.~N.~Wang,
  Phys.\ Rev.\ Lett.\  {\bf 85} (2000) 3591;
  X.~N.~Wang and X.~F.~Guo,
  Nucl.\ Phys.\ A {\bf 696} (2001) 788.

\bibitem{Qiu:1990xy}
  J.~W.~Qiu and G.~Sterman,
  Nucl.\ Phys.\ B {\bf 353} (1991) 137.

\bibitem{Qiu:2002mh}
  J.~W.~Qiu,
  Nucl.\ Phys.\ A {\bf 715} (2003) 309
  [arXiv:nucl-th/0211086].

\bibitem{Qiu:2003cg}
  J.~W.~Qiu,
  in CERN Yellow Report (CERN-2004-009 on ``Hard probes in heavy-ion
  collisions at the LHC'', edited by M. Mangano, H. Satz, and
  U. Wiedemann (Geneva, Switzerland, 2004), PP14 
  [arXiv:hep-ph/0305161].

\bibitem{Qiu:2003vd}
  J.~W.~Qiu and I.~Vitev,
  Phys.\ Rev.\ Lett.\  {\bf 93} (2004) 262301
  [arXiv:hep-ph/0309094].

\bibitem{Jaffe:1981td}
  R.~L.~Jaffe and M.~Soldate,
  Phys.\ Lett.\ B {\bf 105} (1981) 467;
  Phys.\ Rev.\ D {\bf 26} (1982) 49.

\bibitem{Qiu:1990xx}
  J.~W.~Qiu and G.~Sterman,
  Nucl.\ Phys.\ B {\bf 353} (1991) 105.

\bibitem{Qiu:2004da}
  J.~W.~Qiu and I.~Vitev,
  arXiv:hep-ph/0405068.

\bibitem{OPE}
  K.G. Wilson, Phys. Rev. {\bf 179} (1969) 1499.

\bibitem{LQS1} 
  M.~Luo, J.~W.~Qiu and G.~Sterman,
  Phys.\ Lett.\ B {\bf 279} (1992) 377;
  Phys.\ Rev.\ D {\bf 50} (1994) 1951.

\bibitem{Brock:1993sz}
R.~Brock {\it et al.}, 
Rev.\ Mod.\ Phys.\  {\bf 67}, 157 (1995).

\bibitem{Guo:2001tz}
  X.~F.~Guo, J.~W.~Qiu and W.~Zhu,
  Phys.\ Lett.\ B {\bf 523} (2001) 88
  [arXiv:hep-ph/0110038].

\bibitem{LQS} 
  M.~Luo, J.~W.~Qiu and G.~Sterman,
  Phys.\ Rev.\ D {\bf 49} (1994) 4493.

\bibitem{AM}
  A.~H.~Mueller and J.~W.~Qiu,
  Nucl.\ Phys.\ B {\bf 268} (1986) 427.

\bibitem{Pumplin:2002vw}
J.~Pumplin {\em et al.}, 
JHEP {\bf 0207}, 012 (2002).

\bibitem{Qiu:2004qk}
  J.~W.~Qiu and I.~Vitev,
  Phys.\ Lett.\ B {\bf 587} (2004) 52
  [arXiv:hep-ph/0401062].

\bibitem{vitev-dis}
  I.~Vitev,
  arXiv:hep-ph/0506039.

\bibitem{Abe:1998ym}
K.~Abe {\it et al.}, 
Phys.\ Lett.\ B {\bf 452}, 194 (1999).

\bibitem{DISdata}
M.~Arneodo {\it et al.}, 
Nucl.\ Phys.\ B {\bf 441}, 12 (1995);
P.~Amaudruz {\it et al.}, 
Nucl.\ Phys.\ B {\bf 441}, 3 (1995);
M.R.~Adams {\it et al.}, 
Z.\ Phys.\ C {\bf 67}, 403 (1995);
Phys.\ Rev.\ Lett.\  {\bf 68}, 3266 (1992).


\bibitem{GLR}
  L.~V.~Gribov, E.~M.~Levin and M.~G.~Ryskin,
  Phys.\ Rept.\  {\bf 100} (1983) 1.

\bibitem{Eskola-NPB}
  K.~J.~Eskola, H.~Honkanen, V.~J.~Kolhinen, J.~w.~Qiu and C.~A.~Salgado,
  arXiv:hep-ph/0302185.

\bibitem{Zhu:2004zw}
  W.~Zhu,
  Nucl.\ Phys.\ A {\bf 753} (2005) 206
  [arXiv:hep-ph/0408328].

\bibitem{Qiu-Kang}
  J.~W.~Qiu and Z. Kang, in preparation.

\bibitem{NA38NA50-dimuon}
M.C. Abreu {\it et al.}, NA38 and NA50 Collaborations,
Eur. Phys. J. {\bf C14}, 443 (2000).

\bibitem{OpenCharm}
C. Y. Wong and Z. Q. Wang, Phys. Lett. {\bf B 367}, 50 (1996).

\bibitem{thermal-l}
R. Rapp and E. Shuryak, Phys. Lett. {\bf B 473}, 13 (2000).

\bibitem{Meson-meson}
Z. Lin  and X. N. Wang, Phys. Lett. {\bf B 444}, 245 (1998).

\bibitem{Qiu:2001zj}
  J.~W.~Qiu and X.~Zhang,
  Phys.\ Lett.\ B {\bf 525} (2002) 265
  [arXiv:hep-ph/0109210].

\bibitem{Guo:1998rd}
X.~Guo,
Phys.\ Rev.\ D58 (1998) 114033.
[hep-ph/9804234].

\bibitem{E772-dy}
D. M. Alde {\it et al.}, E772 Collaboration, 
Phys. Rev. Lett. {\bf 64}, 2479 (1990).

\bibitem{NA10-dy}
P. Bordale, {\it et al.}, NA 10 Collaboration,
Phys. Lett. {\bf B193} (1987) 373.

\bibitem{E683}  D.\ Naples {\it et al.}\ (E683 Collaboration)
Phys.\ Rev.\ Lett.\ 72 (1994) 2341.

\bibitem{Gyulassy:2002yv}
M.~Gyulassy, P.~Levai and I.~Vitev,
Phys.\ Rev.\ D {\bf 66}, 014005 (2002);
J.~W.~Qiu, I.~Vitev,
Phys.\ Lett. \ B {\bf 570}, 161 (2003);
I.~Vitev,
Phys.\ Lett.\ B {\bf 562}, 36 (2003).

\bibitem{CSS-fac}
  J.~C.~Collins, D.~E.~Soper and G.~Sterman,
  Adv.\ Ser.\ Direct.\ High Energy Phys.\  {\bf 5} (1988) 1
  [arXiv:hep-ph/0409313].

\bibitem{Vitev:2004kd}
  I.~Vitev,
  J.\ Phys.\ G {\bf 31} (2005) S557
  [arXiv:hep-ph/0409297].

\bibitem{Rak:2004gk}
J.~Rak,
J.\ Phys.\ G {\bf 30}, S1309 (2004);
K.~Filimonov,
J.\ Phys.\ G {\bf 30}, S919 (2004).


\bibitem{Adams:2003im}
J.~Adams {\it et al.}, 
Phys.\ Rev.\ Lett.\  {\bf 91}, 072304 (2003).



\bibitem{Ogawa:2004sy}
A.~Ogawa, 
nucl-ex/0408004.


\end{thebibliography}
\end{document}